\documentclass[twocolumn,aps,nofootinbib,prl,showpacs,floatfix]{revtex4}
\usepackage{graphicx, epsfig, array}
\def\be{\begin{equation}}
\def\ee{\end{equation}}
\def\bea{\begin{eqnarray}}
\def\eea{\end{eqnarray}}

\def\reff@jnl#1{{\rm#1\/}}

\def\aj{\reff@jnl{AJ}}                  % Astronomical Journal
\def\araa{\reff@jnl{ARA\&A}}            % Annual Review of Astron and Astrophys
\def\apj{\reff@jnl{ApJ}}                        % Astrophysical Journal
\def\apjl{\reff@jnl{ApJ}}               % Astrophysical Journal, Letters
\def\apjs{\reff@jnl{ApJS}}              % Astrophysical Journal, Supplement
\def\ao{\reff@jnl{Appl.Optics}}         % Applied Optics
\def\apss{\reff@jnl{Ap\&SS}}            % Astrophysics and Space Science
\def\aap{\reff@jnl{A\&A}}               % Astronomy and Astrophysics
\def\aapr{\reff@jnl{A\&A~Rev.}}         % Astronomy and Astrophysics Reviews
\def\aaps{\reff@jnl{A\&AS}}             % Astronomy and Astrophysics, Supplement
\def\azh{\reff@jnl{AZh}}                        % Astronomicheskii Zhurnal
\def\baas{\reff@jnl{BAAS}}              % Bulletin of the AAS
\def\jrasc{\reff@jnl{JRASC}}            % Journal of the RAS of Canada
\def\memras{\reff@jnl{MmRAS}}           % Memoirs of the RAS
\def\mnras{\reff@jnl{MNRAS}}            % Monthly Notices of the RAS
\def\pra{\reff@jnl{Phys.Rev.A}}         % Physical Review A: General Physics
\def\prb{\reff@jnl{Phys.Rev.B}}         % Physical Review B: Solid State
\def\prc{\reff@jnl{Phys.Rev.C}}         % Physical Review C
\def\prd{\reff@jnl{Phys.Rev.D}}         % Physical Review D
\def\prl{\reff@jnl{Phys.Rev.Lett}}      % Physical Review Letters
\def\pasp{\reff@jnl{PASP}}              % Publications of the ASP
\def\pasj{\reff@jnl{PASJ}}              % Publications of the ASJ
\def\qjras{\reff@jnl{QJRAS}}            % Quarterly Journal of the RAS
\def\skytel{\reff@jnl{S\&T}}            % Sky and Telescope
\def\solphys{\reff@jnl{Solar~Phys.}}    % Solar Physics
\def\sovast{\reff@jnl{Soviet~Ast.}}     % Soviet Astronomy
\def\ssr{\reff@jnl{Space~Sci.Rev.}}     % Space Science Reviews
\def\zap{\reff@jnl{ZAp}}                        % Zeitschrift fuer Astrophysik
\def\nat{\reff@jnl{Nature}}             % Nature 

\def\fun#1#2{\lower3.6pt\vbox{\baselineskip0pt\lineskip.9pt
        \ialign{$\mathsurround=0pt#1\hfill##\hfil$\crcr#2\crcr\sim\crcr}}}

\newcommand\eV{\rm eV}

\newcommand\lsim{\mathrel{\rlap{\lower4pt\hbox{\hskip1pt$\sim$}}
    \raise1pt\hbox{$<$}}}
\newcommand\gsim{\mathrel{\rlap{\lower4pt\hbox{\hskip1pt$\sim$}}
    \raise1pt\hbox{$>$}}}

\def\dslash{\not{\hbox{\kern-2pt $\partial$}}}
\def\Dslash{\not{\hbox{\kern-4pt $D$}}}
\def\Oslash{\not{\hbox{\kern-4pt $O$}}}
\def\Qslash{\not{\hbox{\kern-4pt $Q$}}}
\def\pslash{\not{\hbox{\kern-2.3pt $p$}}}
\def\kslash{\not{\hbox{\kern-2.3pt $k$}}}
\def\qslash{\not{\hbox{\kern-2.3pt $q$}}}

 \newtoks\slashfraction
 \slashfraction={.13}
 \def\slash#1{\setbox0\hbox{$ #1 $}
 \setbox0\hbox to \the\slashfraction\wd0{\hss \box0}/\box0 }

\begin{document}

\setlength{\unitlength}{1mm}
%\twocolumn[\hsize\textwidth\columnwidth\hsize\csname@twocolumnfalse\endcsname]
\title{Is cosmology compatible with sterile neutrinos ?}

\author{Scott Dodelson}
\affiliation{Particle Astrohysics Center, FERMILAB, Batavia, IL 60510-0500}
\author{Alessandro Melchiorri}
\affiliation{Physics Department and sezione INFN, University of Rome ``La Sapienza'', Ple Aldo Moro 2, 00185 Rome, Italy}
\author{An\v{z}e Slosar}
\affiliation{Faculty of Mathematics and Physics, University of
  Ljubljana, 1000 Ljubljana, Slovenia}

\date{\today}%
%\maketitle % use with old revtex !

\begin{abstract}
  By combining data from cosmic microwave background (CMB) experiments
  (including the recent WMAP third year results), large scale structure
  (LSS) and Lyman-$\alpha$ forest observations, we constrain the
  hypothesis of a fourth, sterile, massive neutrino.  
  For the 3 massless + 1 massive neutrino case
  we bound the mass of the sterile neutrino to $m_{s}<0.26\eV$ ($0.44\eV$) at
  $95$\% (99.9\%) c.l.. These results exclude at high significance the sterile
 neutrino hypothesis as an explanation of the LSND anomaly. 
  We then generalize the analysis to account for
  active neutrino masses (which tightens the limit to $m_{s}<0.23\eV$ ($0.42\eV$)
  and the possibility that the sterile abundance is not thermal.
  In the latter case, the contraints in the (mass, density) plane
  are non-trivial. For a mass of $>1 \eV$ or $<0.05 \eV$ the
  cosmological energy density in sterile neutrinos is always
  constrained to be $\omega_{\nu} <0.003$ at $95 \%$ c.l.. However,
  for a sterile neutrino mass of $\sim 0.25 \eV$, $\omega_\nu$ can be
  as large as $0.01$.  

\end{abstract}
\bigskip
%\pacs{PACS Numbers: }

\maketitle

{\it Introduction.---} Recent cosmological data coming from measurements of the Cosmic
Microwave Background (CMB) anisotropies (see e.g. 
\cite{2006astro.ph..3451H,2006astro.ph..3450P}), galaxy clustering
(see e.g. \citep{2004PhRvD..69j3501T}) and Lyman-alpha forest clouds
\cite{2005PhRvD..71j3515S} are in a spectacular agreement with the
expectations of the so-called standard model of structure formation,
based on primordial adiabatic inflationary perturbations and a
cosmological constant.

Since the model works so well, the ambitious idea of using cosmology
to test aspects of particle
physics is becoming a reality.  An excellent example of this comes
from the new cosmological constraints on neutrino physics.

Cosmological neutrinos have a profound impact on
cosmology since they change the expansion history of the universe
and affect the growth of perturbations (see
\cite{2004PhRvD..69h3002B} for a detailed account). 
Recent analyses (see e.g. \citep{2005PhRvD..71j3515S,2004PhRvD..70k3003F}) 
have indeed constrained the neutrino mass in the context of 
three-flavour mixing 
to be $m_{\nu} < 0.16$eV ($m_{\nu} <0.45$eV without Lyman-$\alpha$
forest data) with a greater accuracy than laboratory beta decay experiments 
which suggest $m_{\nu} < 2.2$eV (see \cite{2004PhRvD..70k3003F} and references therein).

A possible discrepancy between cosmology and beta decay or neutrino
oscillation experiments might provide valuable information for the
presence of systematics or new physics. At the moment, the claimed
and highly debated detection of a neutrino mass in the range $0.17 \eV <
m_{\beta\beta} < 2.0 \eV$ at $99 \%$ c.l. \citep{Klapdor-Kleingrothaus:2004wj}
from the Heidelberg-Moscow double beta decay experiment is at odds
with the cosmological bound.

While the neutrino masses are very difficult to measure
experimentally, mass differences between neutrino mass eigenstates
($m_1$,$m_2$,$m_3$) have now been measured in oscillation
experiments.  Observations of atmospheric neutrinos suggest a squared
mass difference of $\Delta m^2 \sim 3 \times 10^{-3} \eV^2$, while
solar neutrino observations, together with results from the KamLAND
reactor neutrino experiment, point towards $\Delta m^2 \sim 5 \times
10^{-5} \eV^2$. The two measured mass differences are easily accomodated in 
simple extensions of the
Standard Model by giving masses to at least two of the neutrinos. If these masses
are greater than ~$\sim0.1$ eV, all three neutrinos must be nearly degenerate, with small
differences accounting for the observations.

Results from the Liquid Scintillator Neutrino
Detector (LSND) \citep{Aguilar:2001ty} challenge the simplicity of
this picture.  The LSND
experiment reported a signal for ${\bar \nu}_{\mu} \rightarrow {\bar
  \nu}_{e}$ oscillations in the appearance of ${\bar \nu}_{e}$ in an
originally ${\bar \nu}_{\mu}$ beam.  To reconcile the LSND anomaly with results on neutrino
mixing and masses from atmospheric and solar neutrino oscillation
experiments, one needs additional mass eigenstates.  
One possibility is that these
additional states are related to right-handed neutrinos, for which bare mass terms ($M\nu_R\nu_R$)
are allowed by all symmetries. These would
are {\it sterile}, i.e. not present in
${SU(2)_L}\times{U(1)_\gamma}$ interactions.  The ``$3+1$ sterile''
neutrino explanation assumes that the ${\bar \nu}_{\mu} \rightarrow
{\bar \nu}_{e}$ oscillation goes through ${\bar \nu}_{\mu} \rightarrow
{\bar \nu}_{s} \rightarrow {\bar \nu}_{e}$. The additional sterile
state is separated by the three active states by a mass scale in the
range of $0.6 \eV^2 < \Delta m_{LSND}^2 < 2 \eV^2$. Constraints from long baseline
experiments are threatening this 
interpretation~\cite{Declais:1994su,Stockdale:1984cg,Dydak:1983zq,Apollonio2003,Armbruster:2002mp,Astier:2003gs}; 
it is possible that more than one
sterile neutrino is necessary to explain LSND~\cite{Sorel:2003hf}.
The LSND signal will be soon tested by the MINI-BOONE experiment,
expected to release the first results at the beginning of the next
year.

In the meantime, given the increased quality in the data, it is
timely to test the sterile neutrino hypothesis using
cosmological observations. Several
recent analyses have already provided interesting cosmological
constraints on a fourth massive neutrino
\citep{2005PhRvD..71j3515S,2005PhRvD..71f3534V,2003JCAP...05..004H}.  Here we generalize 
these in several ways: first, while
previous work has considered the case of $3$ (massless) + $1$
(massive) neutrino, here we also allow masses for the standard $3$
neutrinos, as required by oscillation experiments.  Second, we use 
updated cosmological datasets, including the new results from the
WMAP satellite \cite{2006astro.ph..3451H,2006astro.ph..3450P} 
and BOOMERAG-2K2 experiment \citep{2005astro.ph..7503M}.  Finally, the strength 
of the interactions of a neutrino determines its cosmological abundance. Given
how little we know about sterile neutrino interactions (or mass mixing), it therefore
seems reasonable to allow the sterile abundance to be a free parameter. Of course, if a
sterile neutrino can have any abundance (including zero), there is no mass limit. However, we
will see that the constraints in the (mass, density) plane are highly non-trivial.

{\it Cosmology.---} The three active neutrinos interact via the well-measured
weak interactions. These interactions ensure that they were in thermal
equilibrium at early times until they decouple from the primordial plasma
slightly before electron-positron annihilation. After decouple, they maintain
an equilibrium distribution of a massless fermion with a temperature lower than
the photon temperature by a factor of $(4/11)^{1/3}$.
This introduces a
well-known relation between the energy density of the active neutrinos
and their total mass: 
\be
\omega_\nu \equiv {\rho_\nu\over \rho_{\rm cr}} h^2 =  0.0106 {m_\nu\over {\rm eV}}
\ee
where $\rho_{\rm cr}$ is the critical energy density, $h$ parametrizes the Hubble constant via
$H_0=100 h$ km sec$^{-1}$ Mpc $^{-1}$, and here and throughout $m_\nu$ refers to the sum of all {\it active}
neutrino masses. So, for example, if the three neutrinos are nearly dengerate, they each have a mass 
approximately equal to
$m_\nu/3$.

While sterile neutrinos, by defintion, do not have weak interactions, they are not pure mass eigenstates.
As such, oscillations in the early universe can transform the thermal active neutrinos into a sterile
neutrino~\cite{Langacker:1989sv,Barbieri:1990vx}. Thermalization occurs if
\be \Delta m_{\rm LSND}^2 \sin^4\theta > 3\times 10^{-6} {\rm eV}^2
,\ee
where $\theta$ is an effective mixing angle. In the simplest models with one sterile neutrino, 
this condition is satisfied, so 
$\omega_s = 0.0106 (m_s/{\rm eV})$, but there are many ways of evading thermalization~\cite{Abazajian:2002bj}.
Indeed, if one light sterile neutrino exists, there is every reason to expect one or two more and these
considerably complicate the thermalization analysis. 
It is, for example, possible to have super-thermal abundances if a
heavier sterile state decays at relatively late times to a lighter state. In short, if a sterile neutrino
exists, its cosmological density is much more uncertain than that of the active neutrinos.

Sterile neutrinos influence the development of inhomogeneities and anisotropies in the universe by 
changing the epoch of equality and by suppressing perturbations via freestreaming. The epoch at which
the energy density in non-relativistic matter equals that in radiation dictates when structure begins to
grow. This leaves an imprint on the matter power spectrum~\cite{Dodelson:2003ft}: there is a peak at the scale equal to the horizon
at the epoch of equality. If this epoch is close to recombination, the residual radiation 
causes gravitational potentials to
decay, and this time variation produces an
early Intergated Sachs-Wolfe (ISW) effect, boosting the power on scales near the horizon. The main effect
of freestreaming is a suppression of power on scales smaller (wavenumber $k$ larger) than 
\be
k_{\rm fs} = 0.01 \left( {m_s\over {\rm eV}} \right)^{1/2} {\rm Mpc}^{-1}
\ee
with suppression proportional to $\omega_s/\omega_m$, where $\omega_m\equiv \Omega_mh^2$,
and $\Omega_m$ is the total energy density of non-relativistic matter (baryons plus cold dark matter)
in units of the critical density.

\begin{figure}%[!t]
\begin{center}
\includegraphics[width=1.0\linewidth]{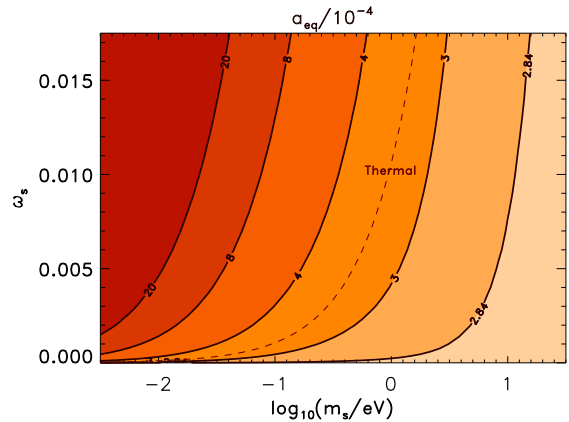}
\caption{The epoch of equality $a_{\rm eq}$ as a function of mass of sterile neutrino
  and its energy density. The non-relativistic matter density here is fixed to $\omega_m=0.15$, so
  that in the standard 3-neutrino model, 
  $a_{\rm EQ}=2.82\times 10^{-4}$. Notice that, at fixed $\omega_s$, $a_{\rm eq}$ rises very rapidly for
  lower masses since the neutrinos behave as
  radiation. Thermalized neutrinos lie along the dashed curve.}
\label{snu}
\end{center}
\end{figure}

In the standard cosmology, with three massless neutrinos, the scale factor at equality is $a_{\rm EQ}=
2.82\times 10^{-4} (0.15/\omega_m)$ . A sterile neutrino is relativistic until its temperature drops beneath its mass, 
so masses of order an eV raise the question: what does it count as, matter or radiation? 
Since the Hubble rate scales as $a^{-2}$ in a radiation
dominated universe and $a^{-3/2}$ in a matter dominated universe, we define the epoch of equality
as the moment when
\begin{equation}
  \label{eq:2}
  \frac{d\ln H}{d \ln a} (a_{\rm eq}) = -\frac{7}{4}
.\end{equation}
This definition agrees well with the standard definition for massless neutrinos.
The dependence of $a_{\rm eq}$  on the sterile neutrino
parameters $m_s$ and $\omega_s$ is plotted in Figure
\ref{snu}. This figure suggests that in the limit of very small $m_{s}$, any
appreciable $\omega_{s}$ will be excluded because neutrinos behave 
essentially as radiation and shift the redshift of matter-radiation
equality significantly, producing an unaccetably large ISW effect.

The amount of suppression due to freestreaming increases as the
density increases (from top to bottom in Fig.~\ref{snu}), but the
large scales (from which constraints derive) cease to be affected as
the neutrino mass increases (from left to right). Therefore, at fixed
$\omega_s$, constraints from freestreaming are tighter for {\it small}
neutrino masses. Note that this differs from the thermal case (dashed
curve in Fig.~\ref{snu}). In that case, the neutrino density increases
with the mass, so there is {\it more} suppression at high masses.

\begin{figure}%[!t]
\begin{center}
\includegraphics[width=1.0\linewidth]{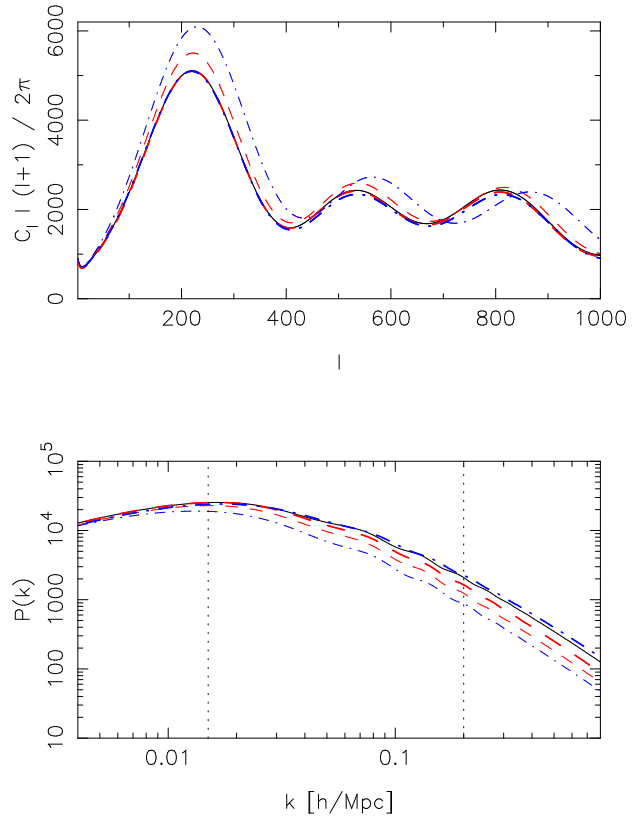}
\caption{Effect of extra sterile neutrino on the CMB (top) and LSS
(bottom) power spectra. Thin lines correspond to standard model, sterile
neutrino of mass $m=1$eV (dashed) $m=0.3$eV (dot-dashed) and fixed
sterile density $\omega_{s}=0.01$. These curves are normalised to
large scale $C_\ell$. Thick dashed and dot-dashed curves
correspond to models, which in addition to having sterile mass have
had dark matter density increased to match standard $a_{\rm eq}$ and
$h$ increased to match CMB peak positions and were normalised at the first
peak. Dotted vertical lines on the bottom plot enclose  the area
where LSS experiments are currently sensitive to with thick line
normalisations chosen to illustrate the fact that the $1$eV model is a
poorer fit than $0.3$eV model. See text for discussion. \label{bigure3}}
\end{center}
\end{figure}

{\it Data analysis and Results-- } 
To obtain contraints on sterile neutrino parameters, we use the
publicly available Markov Chain Monte Carlo (MCMC) package \texttt{CosmoMC}~\citep{2002PhRvD..66j3511L}. The linear
perturbations engine CAMB \cite{2000ApJ...538..473L} of the software has been generalized in
several ways. First, we allow for a non-thermal sterile neutrino density.  
Second,  we allow for the possibility that the active neutrinos have mass different than the
sterile neutrino.

In the MCMC, we sample the following $8$ dimensional set
of cosmological parameters, adopting flat priors on them: The log mass of thermal 
sterile neutrinos $\log m_{s}$ and $\omega_{\nu}$, the energy
density of $3$ degenerate standard massive neutrinos
$\omega_{\nu}=m_{\nu}/(94.1 \eV)$, the physical baryon and CDM
densities, $\omega_b=\Omega_bh^2$ and $\omega_c=\Omega_ch^2$, the
ratio of the sound horizon to the angular diameter distance at
decoupling, $\Theta_s$, the scalar spectral index and the overall
normalisation of the spectrum, $n_s$ and $A_s$, and, finally, the
optical depth to reionisation, $\tau_r$. We consider
purely adiabatic initial conditions, impose flatness, and do not
include gravitational waves.

We include the WMAP three-year data
\cite{2006astro.ph..3451H,2006astro.ph..3450P} (temperature and
polarisation) with the routine for computing the likelihood supplied
by the WMAP team \cite{2006astro.ph..3449S}, as well as the CBI
\citep{2004ApJ...609..498R}, VSA \cite{2004MNRAS.353..732D}, ACBAR
\cite{2002AAS...20114004K} and BOOMERANG-2k2
\cite{2005astro.ph..7503M} measurements of the CMB on scales
smaller than those sampled by WMAP. In addition to
the CMB data, we also consider the constraints on the real-space power
spectrum of galaxies from the SLOAN galaxy redshift survey (SDSS)
\cite{2004ApJ...606..702T} and the 2dF galaxy redshift survey 
\cite{2005MNRAS.362..505C} and
Lyman-alpha forest clouds \cite{McDonald:2004eu,McDonald:2004xn} from
the SDSS, the gold sample of the recent supernova type Ia data
\cite{2004ApJ...607..665R}, the latest SNLS supernovae
data \cite{2006A&A...447...31A} and the constraints from the baryonic
acoustic oscillations detected in the Luminous Red Galaxies sample of
the SDSS \cite{2005ApJ...633..560E} \footnote{There is a negligible overlap 
between the constraints from the 2dFGRS, SDSS and SDSS LRG analysis, 
as there are galaxies in common in all three data sets.}

The details of the analysis are
the same as those in \cite{Seljak:2006bg} and the reader is invited to
check that paper to examine what constraints the above datasets give
for other models including standard 3 degenerate massive neutrinos
case.

If the active neutrino masses are fixed to zero and the sterile
neutrino abundance is thermal (similar to the assumptions imposed in
Ref.~\cite{2005PhRvD..71j3515S}), the upper limit on the sterile
neutrino mass is $0.26$eV ($0.44$eV) (all at 95\% (99.9\%) c.l.). Of
course the active neutrino masses are not zero. Taking them as a free
parameter leads to an upper limit on the sterile neutrino mass of
$0.23$eV ($0.42$eV). This is marginally tighter than the $m_\nu=0$
constraint because the limit is really on the sum of all neutrino
masses. Fixing the active masses to zero allows the maximum $m_s$.
Relaxing this restriction leaves less room for a large $m_s$. We have
found some sensitivity to the mass {\it difference} of the sterile and
active states (and this might be measurable with future data), but
current data really constrain only the sum of all neutrino masses.

\begin{figure}%[!t]
\begin{center}
\includegraphics[angle=-90,width=0.8\linewidth]{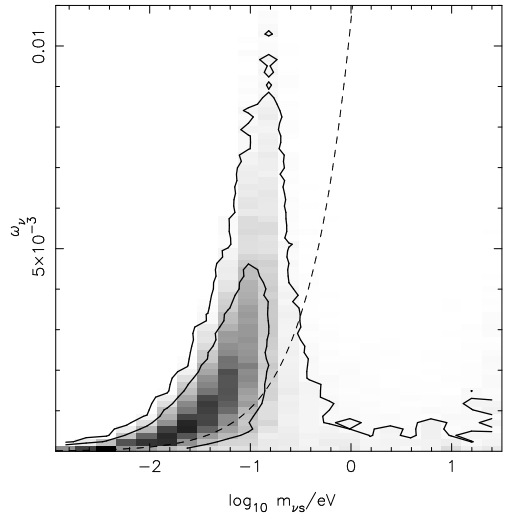}
\caption{$1$,$2$-$\sigma$ constraints on the sterile neutrino mass and abundance.} \label{num}
\end{center}
\end{figure}

We now generalize further and allow the sterile neutrino abundance
$\omega_s$ to vary.  Fig.~\ref{num} shows the constraints in the
$\omega_{s}$-$m_{s}$ plane.  Note the distinct peak around the region
of $m_{s} \sim 0.25 {\rm eV}$, presenting an allowed region of
parameter space with anomalously large values of $\omega_{s}$.  To the
left of this peak, $a_{\rm EQ}$ is very large and the resulting ISW
effect precludes agreement with CMB data. When $m_s$ is in the allowed
regime, $a_{\rm EQ}$ would still be too large were $\omega_{m}$
fixed. However, a model with larger $\omega_m (\sim 0.18)$ leads to an
even smaller, acceptable $a_{\rm EQ}$. Fortuitously, the enhanced cold
matter density also mitigates the freestreaming suppression (which
scales as $\omega_c^{-1}$). At larger neutrino mass ($\sim 1$eV),
additional cold matter would make $a_{\rm EQ}$ too {\rm small}, so
$\omega_m$ must be closer to $0.13$ and the free-streaming suppression
becomes relevant again, preventing agreement with large scale
structure. This is illustrated in Fig.~\ref{bigure3}. Here we show the
angular CMB anisotropy and matter power spectrum for different masses
at fixed $\omega_s$. The suppression due to free-streaming is evident
in the power spectrum and clearly becomes more severe for smaller
masses. However, increasing dark matter density to match the epoch of
matter-radiation equality opposes this effect. Crucial to this
interpretation is the realization that the matter-radiation equality
is very thoroughly measured by the present-day experiments with little
model-dependence. The constraint can be summarised in $a_{\rm eq}\sim
(2.95 \pm 0.13) \times 10^{-4}$.

{\it Conclusions--} By combining data from cosmic microwave background
experiments, galaxy clustering and Ly-alpha forest observations we
have constrained the hypothesis of a fourth, sterile, massive
neutrino, as an explanation of the LSND anomaly.  For the 3 massless +
1 massive thermal neutrino case we bound the mass of the sterile
neutrino to $m_{\nu}<0.26\eV$($0.44\eV$) at 95 (99.9) \% c.l.  Marginalizing over
active neutrino masses improves the limit to $m_{\nu}<0.23\eV$($0.42\eV$).  These
limits are incompatible at more than $3 \sigma$ with the LSND result
$0.6\eV^2 < \Delta m_{\rm LSND}^2 < 2 \eV^2$ ($95 \%$ C.L.).
Moreover, our analysis renders the LSND anomaly incompatible at high
significance with a degenerate active neutrino scenario and viceversa.
If we allow for the possibility of a non-thermal sterile neutrino, we
find that the upper limit of allowed energy density in the sterile
neutrino is a strong function of mass. In particular, for $m_s<1 \eV$
or $>0.05 \eV$ the cosmological energy density in sterile neutrinos is
always constrained to be $\omega_{s} <0.003$, but that for sterile
neutrino mass of $\sim 0.25\eV$, $\omega_s$ can be as large as
$0.01$eV.

The results presented in this paper rely on the assumption that
systematics in the public datasets we analyzed (WMAP, Lyman-$\alpha$,
etc.)  are under control. We argue that this is likely: the datasets
are large enough that detailed systematics checks -- e.g. dividing the
data into multiple subsets, constructing {\it quiet} channels that
should see nothing, and cross-correlating different bands to reduce
noise -- have been performed.  We also checked that if we drop either
small scale CMB, LSS or Lyman-$\alpha$ dataset from the analysis, the
constraints simply weaken without any systematic change in the
results.

The results presented here also rely on the assumption of a
theoretical cosmological model based on a large but limited set of
parameters.  Extensions of the parameter space -- e.g., inclusion of
isocurvature modes, gravity waves or a different parametrization for
the dark energy component -- may modify our conclusions. Those
modifications are however not needed by current data and some of them
may well lead to stronger limits. Indeed the simple cosmological model with only
a handful of parameters does an excellent job explaining a wide variety of data. If the LSND anomaly is
confirmed by MINIBOONE, we will have been proved wrong, and cosmologists will 
need to re-examine the entire framework on which these very tight constraints rest.

\textit{Acknowledgements} SD is supported by the DOE and by NASA grant
NAG5-10842.  AM is supported by MURST through COFIN contract
no. 2004027755. AS is supported by the Slovene Ministry of Higher
Education. We acknowledge useful discussions with Antony Lewis,
Patrick McDonald, Stephen Parke and Uro\v{s} Seljak. We also thank
Patrick McDonald for the public Lyman $\alpha$ likelihood code that
was used in this work. AS additionaly thanks Oxford Astrophysics for
hospitality during the visit during which part of this project has
been completed.

\bibliography{../BibTeX/cosmo,../BibTeX/cosmo_preprints,scott}

\end{document}